\documentstyle[epsf,aps,eqsecnum,preprint,12pt]{revtex}

\newcommand{\teb}[1]{{\displaystyle\oalign{$#1$\crcr\hidewidth\vbox 
                     to.2ex{\hbox{\char126}\vss}\hidewidth}}}
\newcommand{\bigfrac}[2]{\displaystyle{\frac{#1}{#2}}}
\newcommand{\bigint}{\displaystyle{\int}}
\begin{document}
\draft
%\preprint{HEP/123-qed}
\title{The structure of the graviton self-energy at finite temperature}

\author{F. T. Brandt and J. Frenkel}

\address{Instituto de F\'\i sica, Universidade de S\~ao Paulo, \\
S\~ao Paulo, 01498 SP, Brasil}

\date{\today}
\maketitle
\begin{abstract}
We study the graviton self-energy function in a general gauge,
using a hard thermal loop expansion which includes
terms proportional to $T^4$, $T^2$ and $\log(T)$.
We verify explicitly the gauge independence of the leading $T^4$
term and 
obtain a compact expression for the sub-leading $T^2$ contribution.
It is shown that the logarithmic term has the same structure as 
the ultraviolet pole part of the $T=0$ self-energy
function. We argue that the gauge-dependent part of the $T^2$ 
contribution is effectively canceled in the dispersion relations 
of the graviton plasma, and present the solutions of these
equations.
\end{abstract}

\section{Introduction}

When the temperature $T$ is high compared with the typical momentum scale 
but well below the Planck scale, all the n-graviton thermal
Green functions can be computed in the one-loop approximation
using the hard thermal loop expansion. There have been many investigations
where this approach has been employed 
\cite{GroPerGribDon,GriboskyDonoghue,Kobes,Rebhan,FrenTay,Almeida,Kraemmer}. 
An important property which is now well established is the gauge invariance
of the leading high temperature contributions of all n-graviton thermal Green 
functions. The explicit from of these contributions can be obtained
using the equivalence which exists between the formalism of 
Boltzmann transport equation and the high temperature limit of the
thermal Green functions in field theory\cite{Kelly}. Using this
approach (which is explicitly gauge invariant) one can easily show 
that the leading part of all n-point one-loop thermal Green functions is proportional to 
$T^4$ \cite{BrandtFrenkelTaylor}.
These results have also been obtained by standard Feynman diagrammatic calculation
in the Feynman-DeDonder gauge for the one- and two-graviton functions \cite{Rebhan}
as well as for the three-graviton function \cite{BrandtFrenkel}.

One of the interesting physical applications of the 
one- and two-graviton functions is the study of
the dispersion relations\cite{Rebhan} which follow from 
{linear response theory}\cite{LeBellac}.
Since the relevant physical quantities are obtained from 
the {\it poles} \/of the propagator, it is important to verify 
the gauge independence of this procedure. While this is automatically
fulfilled by the leading high temperature contributions, the
inherently gauge dependent {\it sub-leading contributions} \/
to the thermal Green functions require a more detailed investigation.
A similar situation occurs in the case the Yang-Mills theory where
it is known that a gauge independent set of dispersion relations can 
be obtained from the gauge dependent thermal two-gluon function
\cite{Kobes,Weldon}. As far as we know, in contrast with the case of the Yang-Mills theory, 
a gauge independence proof of the 
dispersion relations in quantum gravity beyond the leading order is still missing.

The purpose of the present paper is to investigate this problem
in the case of gravity using the standard Feynman diagrammatic
approach. We will compute  the 1- and 2-graviton functions up to sub-leading
contributions, in a class of general gauges. 
(Some intrinsically gauge independent sub-leading contributions 
have been considered before using the theory 
of a scalar field in interaction with gravity\cite{Almeida}.)
We employ the {\it imaginary time formalism}\cite{Kapusta} 
and express the one-loop thermal Green functions in terms of on-shell 
forward scattering amplitudes (the ``Barton amplitude'')\cite{Barton},
properly generalized in order to account for the quadratic
denominators which arises in the free graviton propagator
when a general gauge fixing term is employed
\cite{BartonGen} \/(see also appendix~\ref{appA}). 
This approach enable us to explore some of the general properties
of the {\it exact} \/graviton self-energy without having to carry out 
explicitly the non-trivial spatial momentum integrations. 
It is also much more straightforward 
to perform the hard thermal loop expansion  
when we start from the forward scattering amplitudes.
Using this approach we were able to obtain
explicit results for the $T^2$ and logarithmic contributions to the 
graviton self-energy. 

Unlike the leading high temperature terms, 
for which the gauge independence is confirmed by our calculation, 
the sub-leading contributions are gauge
dependent. These contributions will 
be employed in the study of the dispersion relations for the transverse and traceless
gravitational modes\cite{Rebhan}.

This paper is organized as follows: In Sect.~\ref{sect2} \/we present the
Lagrangian and the basic definition of the graviton field from which
the Feynman rules are derived. We also
discuss the identities which follows from the gauge invariance of the
theory. In Sect.~\ref{sect3} \/the main results of the calculation of the
one- and two-graviton functions up to the logarithmic contributions
is presented. A very compact expression for the $T^2$ contribution
is obtained and we verify that the logarithmic contribution
is proportional to the ultraviolet pole part of the $T=0$ two-graviton function.
We also derive the general transformation of the
two-graviton function under a change of graviton representation.
In Sect.~\ref{sect4} \/we verify explicitly that the
gauge dependence of the sub-leading $T^2$ contributions to the two-graviton function
is effectively canceled in the dispersion relations.
We then  present the solutions of these equations, which include corrections
of order $T^2$ to the leading $T^4$ contributions, describing the
physical modes for the propagation of waves in a graviton plasma.

\section{Feynman rules and identities}
\label{sect2}

The Feynman rules for the graviton propagator and
self-interactions vertices are obtained from
the following underlying Lagrangian
\begin{equation}
  \label{lagrangian}
  {\cal L}=\frac{2}{\kappa^2}\sqrt{-g}R + 
\frac{1}{\kappa^2\,\xi}\eta_{\mu\nu}
\left(\partial_\rho \sqrt{-g}g^{\rho\mu}\right)
\left(\partial_\sigma \sqrt{-g}g^{\sigma\nu}\right)+
\partial_\mu\chi_\nu\frac{\delta
\sqrt{-g}g^{\mu\nu}}{\delta\epsilon^\lambda}
\eta^\lambda ;\;
\kappa\equiv\sqrt{32\pi G},
\end{equation}
where $R$ is the Ricci scalar, G is the Newton constant and
the parameter $\xi$ defines a family of gauges.
($\xi=1$ is the Feynman gauge and $\xi=0$ is the Landau gauge).
The quantities $\chi_\nu$ and $\eta^\lambda$ are the {\it ghost fields}
\/and the function $\epsilon(x)$ is the infinitesimal generator of
coordinate (gauge) transformations
\begin{equation}\label{transf}
x^\mu\rightarrow x^\mu+\epsilon^\mu(x).
\end{equation}
The calculations in quantum gravity are conveniently performed
using the {\it graviton field} \/$h^{\mu\nu}$ defined 
in terms of the tensor $g^{\mu\nu}$ as
\begin{equation}
  \label{field}
  \sqrt{-g}\,g^{\mu\nu} \equiv \eta^{\mu\nu} + \kappa h^{\mu\nu},
\end{equation}
where $\eta^{\mu\nu}$ is the Minkowski metric.

The Feynman rules can be obtained in a straightforward way
substituting (\ref{field}) into (\ref{lagrangian}) and performing
a perturbative expansion in $\kappa$. 
The 0th order terms are quadratic
in the graviton field and yield the following expression for the graviton 
propagator
\begin{equation}\label{propag}
\begin{array}{lll}
 {\cal D}^{(0)}_{\mu_1\nu_1\mu_2\nu_2}(k)&=&
\bigfrac{-1}{2k^2}\left\{\eta_{\mu_1\mu_2}\eta_{\nu_1\nu_2}+
                           \eta_{\nu_1\mu_2}\eta_{\mu_1\nu_2}-
                           \eta_{\mu_1\nu_1}\eta_{\mu_2\nu_2}\right. \\
&{}& \\ 
&{}&
+\bigfrac{\left(1-\xi\right)}{\left(k^2\right)^2}\left(
2 k_{\mu_1} k_{\nu_1}\eta_{\mu_2\nu_2}+
2 k_{\mu_2} k_{\nu_2}\eta_{\mu_1\nu_1}-
\eta_{\mu_1\nu_1}\eta_{\mu_2\nu_2}\,k^2 
\right. \\
&{}& \\ 
&{}&
\left. \left. 
-k_{\mu_1} k_{\mu_2}\eta_{\nu_1\nu_2}-
k_{\nu_1} k_{\mu_2}\eta_{\mu_1\nu_2}-
k_{\mu_1} k_{\nu_2}\eta_{\nu_1\mu_2}-
k_{\nu_1} k_{\nu_2}\eta_{\mu_1\mu_2}
\right)\right\}.
\end{array}
\end{equation}
In the appendix~\ref{appB} \/we give all the other relevant Feynman rules
employed in this work. 

The choice of 
the graviton field parametrization given by (\ref{field}) 
restricts the gauge parameter dependence
only to the propagator (\ref{propag}), since in this case the second term in
(\ref{lagrangian}) is exactly quadratic in the graviton field $h$.
This is similar to the general {\it linear gauges} \/in Yang-Mills theories. 
Therefore, the gauge dependence of the Green functions computed from these
Feynman rules can be traced back to Eq. (\ref{propag}).

The leading high temperature contributions of all
one-particle irreducible thermal Green functions are 
related to each other through tree-like Ward identities
in the same way as the basic tree vertices \cite{Kobes,FrenTay,BrandtFrenkelTaylor}.
These hard thermal loop identities have been verified for both Yang-Mills theories
and gravity and generalized to any gauge theory whose generators form
a closed algebra\cite{Kobes}.
For our present purposes it will be sufficient to consider 
the identity involving the two-point function.
A simple example is provided by the following tree Ward identity, 
arising from the invariance of the pure Einstein action
$S_{G}\equiv 2/\kappa^2 \int {\rm d}^4 x \sqrt{-g}R$ 
under the transformation given by equation (\ref{transf}),
\begin{equation}\label{X0S0}
X^{(0)}_{\mu_1\nu_1\;\lambda}(k)
S^{(0)\mu_1\nu_1\;\mu_2\nu_2}(k)=0,
\end{equation}
where 
\begin{equation}
  \label{X0}
  X^{(0)}_{\mu\nu\;\lambda}(k)=
k_{\mu}\eta_{\nu\lambda}+k_{\nu}\eta_{\mu\lambda}-k_{\lambda}\eta_{\mu\nu},
\end{equation}
is the tensor generated from the transformation of the graviton
field under (\ref{transf}) and
\begin{equation}
  \label{S0}
\begin{array}{lll}
  S^{(0)}_{\mu_1\nu_1\;\mu_2\nu_2}(k)&=&
-\bigfrac{k^2}{2}\left(\eta_{\mu_1\mu_2}\eta_{\nu_1\nu_2}+
                       \eta_{\mu_1\nu_2}\eta_{\nu_1\mu_2}-
                       \eta_{\mu_1\nu_1}\eta_{\mu_2\nu_2}\right) \\
{}&{}&
+\bigfrac{1}{2}\left(k_{\mu_1}k_{\mu_2}\eta_{\nu_1\nu_2}+
                    k_{\mu_1}k_{\nu_2}\eta_{\nu_1\mu_2}+
                    k_{\nu_1}k_{\nu_2}\eta_{\mu_1\mu_2}+
                    k_{\nu_1}k_{\mu_2}\eta_{\mu_1\nu_2}\right)                 
\end{array}
\end{equation}
comes from the quadratic term in the action without the gauge fixing
term (it is the inverse of the propagator in the limit 
$\xi\rightarrow\infty$).

The tree-like identity which holds for the high temperature limit
of the two-graviton function would be identical to Eq. (\ref{X0S0})
if the one-graviton function (the tadpole), shown in the diagrams in
Fig. 1, were zero. 
The modification introduced by the tadpole changes the right hand side of 
Eq. (\ref{X0S0}) to a non zero quantity when
$S^{(0)}_{\mu_1\nu_1\;\mu_2\nu_2}(k)$ is replaced by the leading
high temperature contribution of $\Pi^{\mu_1\nu_1\;\mu_2\nu_2}(k,u)$,
given by the diagrams in \hbox{Fig. 2} \/
($u$ is a time-like normalized four-vector 
representing the local rest frame of the plasma). This contrasts with the
analogous situation in the case of Yang-Mills 
theories where the anti-symmetry of the group structure constants
trivially makes the tadpole to vanish.
As a consequence of the non-vanishing tadpole,
the general BRST identities will not hold for
the {\it exact} \/finite temperature graviton self-energy.
However, as we will see in the next section, the tadpole diagrams can be 
computed {\it exactly}, \/yielding a result proportional to $T^4$.
Therefore, if we split $\Pi^{\mu_1\nu_1\;\mu_2\nu_2}(k,u)$ as
\begin{equation}
   \label{break}
\Pi^{\mu_1\nu_1\;\mu_2\nu_2}(k,u)=
\Pi^{\mu_1\nu_1\;\mu_2\nu_2}_{leading}(k,u)+
\Pi^{\mu_1\nu_1\;\mu_2\nu_2}_{sub}(k,u),
\end{equation}
the BRST identities derived in the appendix~\ref{appC} \/will
hold for the sub-leading contributions $\Pi^{\mu_1\nu_1\;\mu_2\nu_2}_{sub}(k,u)$,
so that the following identity is satisfied
\begin{equation}
  \label{tHooft}
 X^{(0)}_{\mu_1\nu_1\;\lambda}(k)
\Pi^{\mu_1\nu_1\;\mu_2\nu_2}_{sub}(k,u)
 X^{(0)}_{\mu_2\nu_2\;\delta}(k)=0.
\end{equation}
This identity is analogous to $k_\mu k_\nu \Pi_{\mu\nu}^{QCD}=0$,
where $\Pi_{\mu\nu}^{QCD}$ is the exact gluon self-energy\cite{Weldon}.
Since all the gauge parameter dependence is restricted to the
sub-leading contributions, these identities have an important
r\^ole in the cancellation of the gauge dependence in the
dispersion relations.

\section{The one- and two-graviton functions in a general gauge}
\label{sect3}

In this section we will present the details of the
calculation of the one- and two-graviton functions.
Let us first consider the contributions from the two
tadpole diagrams in figure 1. 
The most involved diagram is the one shown in figure 1a, 
since both the 3-graviton vertex and the general
gauge propagator are involved. Using Eq. (\ref{3grav}) and the
propagator (\ref{propag}), the straightforward contraction of indices 
yields a result which is independent of the parameter 
$\left(1-\xi\right)$. Therefore,
the resulting expression is identical to what is
obtained in the Feynman-DeDonder gauge involving only
the usual {\it quadratic denominators}.
Using the Eq. (\ref{genint}) in
the simple case when ${i}=1$ and ${j}=0$ the following 
result for the one-graviton function is readily obtained
\begin{equation}
  \label{tadpole}
\begin{array}{lll}
 \Gamma_{\mu\nu} & = &
- \kappa \bigfrac{1}{(2\pi)^3}\bigint_0^\infty 
\bigfrac{|\teb{q}|^2{\rm d}|\teb{q}|}
{2 |\teb{q}|}\bigfrac{1}{e^{|\teb{q}|\,Q\cdot u/T}-1} 
\bigint{\rm d}\Omega \; 2\, Q_\mu Q_\nu \\
{} & {} & \\
{}& = & - \kappa \rho \bigint\bigfrac{{\rm d}\Omega}{4\pi} Q_\mu Q_\nu 
=\kappa\bigfrac{\rho}{3}\left(\eta_{\mu\nu}-4 u_\mu u_\nu\right)
  ; \;\;\; \rho\equiv\frac{\pi^2}{30} T^4 
\end{array},
\end{equation}
where $\int{\rm d}\Omega$ denotes the angular integral and 
the four vector $Q_\mu$ is on-shell with components
given by $Q_\mu = (1, \teb{q} / |\teb{q}|)$.

The diagrams contributing to $\Pi^{\mu_1\nu_1\;\mu_2\nu_2}(k,u)$
are shown in figure 2. The contributions associated with each of these
diagrams will involve integrals like the one shown in
Eq. (\ref{genint6}). From the structure of the graviton propagator
given by Eq. (\ref{propag}) we can see that the diagram in figure 2a
is such that each of its terms will involve integrals with 
${i},\,{j}=1,2$, while in the case of the diagram shown in figure 2b, 
all the corresponding integrals have the form of the first term of 
Eq. (\ref{genint6}) with ${j}=0$ and ${i}=1,2$. 
In the case of the ghost loop diagram shown in figure 2c, 
all the terms will involve integrals with ${i}={j}=1$.
Let us first consider the leading high temperature behavior
of these integrals. In this limit, we can perform a hard thermal
loop expansion of the integrand such that the terms with
${i},{j} > 1$ will all be sub-leading. For the terms with
${i},{j} = 1$ we use expansions like
\begin{equation}
  \label{expansion}
  \left.\frac{1}{(q+k)^2}\right |_{q^2=0}=
\frac 1 2 \frac{1}{{q}\cdot k}-
\frac 1 4 \frac{k^2}{\left({q}\cdot k\right)^2}+
\frac 1 8 \frac{\left(k^2\right)^2}
{\left({q}\cdot k\right)^3} + \cdots,
\end{equation}
The case ${i}=1,\,{j} = 0$ (from the diagram in figure 2b) is 
similar to the tadpole diagram giving an exact $T^4$ contribution.
In this way, we obtain the following result for the leading
behavior of the graviton self-energy
\begin{equation}
  \label{PiT4}
\begin{array}{lll}
\left.\Pi_{\mu_1\nu_1\mu_2\nu_2}^{leading}(k,u)\right. & = &
\kappa^2\bigfrac{\rho}{2}\bigint\frac{{\rm d}\Omega}{4\pi}\left(
\bigfrac{k_{\mu_1} Q_{\nu_1} Q_{\mu_2} Q_{\nu_2}}{k\cdot Q}+
\bigfrac{k_{\nu_1} Q_{\mu_1} Q_{\mu_2} Q_{\nu_2}}{k\cdot Q}+ \right. \\
{} & {} & \\
{} & {} & \left.
\bigfrac{k_{\mu_2} Q_{\nu_1} Q_{\mu_1} Q_{\nu_2}}{k\cdot Q}+
\bigfrac{k_{\nu_2} Q_{\nu_1} Q_{\mu_2} Q_{\mu_1}}{k\cdot Q}-
\bigfrac{k^2\,Q_{\mu_1} Q_{\nu_1} Q_{\mu_2} Q_{\nu_2}}{(k\cdot Q)^2}
\right)
\end{array},
\end{equation}
It is worth mentioning that though a na\"{\i}ve power counting would allow for 
a gauge parameter dependence from the third term in the second line of 
Eq. (\ref{propag}), the final result (\ref{PiT4}) is gauge independent 
as one would expect on more general grounds \cite{Kobes}.
Combining the Eqs. (\ref{break}), (\ref{tHooft}) and (\ref{PiT4}) we
obtain the following identity for the exact self-energy
\begin{equation}
  \label{tHooft1}
 X^{(0)}_{\mu_1\nu_1\;\lambda}(k)
\Pi^{\mu_1\nu_1\;\mu_2\nu_2}(k,u)
 X^{(0)}_{\mu_2\nu_2\;\delta}(k)=
2k^2 \,\kappa^2\, \rho \bigint\frac{{\rm d}\Omega}{4\pi}
Q_\lambda Q_\delta=-2k^2 \,\kappa\,\Gamma_{\lambda\delta},
\end{equation}
where in the last term we have used (\ref{tadpole}).
Since the integrand in the right hand side of
the above expression is an elementary expression without
denominators, the same should be true for its left hand side,
up to terms which would vanish after integration. Our calculation
shows that, in fact, the expression obtained from the diagrams in 
figure 2 is such that the exact integrand of $X^{(0)}_{\mu_1\nu_1\;\lambda}
\Pi^{\mu_1\nu_1\;\mu_2\nu_2}X^{(0)}_{\mu_2\nu_2\;\delta}$
does not involve any denominators, being 
identical to the integrand in the right hand side of (\ref{tHooft1}).

We have extended the hard thermal loop expansion in order
to obtain the $T^2$ and the {\it logarithmic contributions}, \/which are
yielded respectively by the terms of degree $-1$ and $-3$ in $|\teb{q}|$
(terms of degree $-2$ in $|\teb{q}|$ are absent due to the symmetry 
$q\leftrightarrow -q$) from the expansion of the integrand in
expressions like (\ref{genint6}) in the region of 
large values of $|\teb{q}|$. After a long
computation we have been able to find the following compact expression
for the $T^2$ contribution
\begin{equation}
  \label{PiT2}
\begin{array}{lll}
\Pi_{\mu_1\nu_1\mu_2\nu_2}^{T^2}(k,u) & = & 
\bigfrac{\kappa^2\, T^2}{12} \left\{4 S^{(0)}_{\mu_1\nu_1\mu_2\nu_2}
+S^{(0)}_{\mu_1\nu_1\rho\sigma}I_F^{\rho\sigma\lambda\delta}
 S^{(0)}_{\lambda\delta\mu_2\nu_2}
\right. \\ {}&{}&
\left. +(1-\xi)\left[S^{(0)}_{\mu_1\nu_1\mu_2\nu_2}
+S^{(0)}_{\mu_1\nu_1\rho\sigma}I_G^{\rho\sigma\lambda\delta}
 S^{(0)}_{\lambda\delta\mu_2\nu_2}
\right]\right\},
\end{array}
\end{equation}
where 
\begin{equation}
      \label{IF}
I_F^{\rho\sigma\lambda\delta}\equiv
\bigint\frac{{\rm d}\Omega}{4\pi}\left[
\frac{4 Q^\rho    Q^\sigma   \eta^{\lambda\delta}
     +4 Q^\lambda Q^\delta   \eta^{\rho\sigma} 
  - 8 Q^\rho    Q^\lambda   \eta^{\sigma\delta}}{(k\cdot Q)^2} 
 - \frac{k^2}{2}
  \frac{Q^\rho Q^\sigma Q^\lambda Q^\delta}{(k\cdot Q)^4}\right]
\end{equation}
and
\begin{equation}
      \label{IG}
I_G^{\rho\sigma\lambda\delta}\equiv
\bigint\frac{{\rm d}\Omega}{4\pi}\left[
\frac{Q^\rho    Q^\sigma   \eta^{\lambda\delta}
     +Q^\lambda Q^\delta   \eta^{\rho\sigma}}{(k\cdot Q)^2} 
\right].
\end{equation}
Using Eq. (\ref{X0S0}) and the structure of (\ref{PiT2}) we immediately
conclude that 
\begin{equation}\label{X0PiT2}
X^{(0)\,\mu_1\nu_1\;\lambda}(k)
\Pi_{\mu_1\nu_1\mu_2\nu_2}^{T^2}(k,u)=0.
\end{equation}
This result can be understood in the context of the BRST identities,
using the results of Appendix~\ref{appC}. It is remarkable that though
this $T^2$ contribution is gauge dependent, it is
transversal to $X^{(0)\,\mu_1\nu_1\;\lambda}$.

It is straightforward to obtain the explicit results for the 
angular integrals in Eqs. (\ref{PiT4}), (\ref{IF}) and (\ref{IG})
in terms of a tensor basis such as the one shown in the table~\ref{tab1}.
Using the following decomposition
\begin{equation}\label{decomposition}
\int{\rm d}\Omega f^{\mu_1\nu_1\mu_2\nu_2}(k,Q) = \sum_{i=1}^{14} c_i(k,u)
T_i^{\mu_1\nu_1\mu_2\nu_2},
\end{equation}
where $f^{\mu_1\nu_1\mu_2\nu_2}(k,Q)$ is a function of degree 2 or
0 in $Q$ respectively for the leading $T^4$ or the $T^2$ contributions,
the coefficients $c_i(k,u)$ are obtained contracting 
both sides of Eq. (\ref{decomposition}) with each of the 14 tensors
of table~\ref{tab1}. The solution of the resulting linear system of 14 equations 
is given in terms of integrals like 
$\int{\rm d}\Omega \left(k\cdot Q\right)^r$, which can be easily evaluated.

In the case of the logarithmic contributions the resulting 
angular integrals $\int{\rm d}\Omega$ can all be parametrized in a 
{\it Lorentz covariant} \/way in terms of the 5 tensors
$T_1^{\mu_1\nu_1\mu_2\nu_2}$, 
$T_4^{\mu_1\nu_1\mu_2\nu_2}$, 
$T_8^{\mu_1\nu_1\mu_2\nu_2}$,
$T_{12}^{\mu_1\nu_1\mu_2\nu_2}$,   
$T_{13}^{\mu_1\nu_1\mu_2\nu_2}$.
The result can be expressed in terms of the 
$T=0$ graviton self-energy \cite{Delbourgo} in the following way
\begin{equation}
  \label{Pilog}
\Pi_{\mu_1\nu_1\mu_2\nu_2}^{log}(k,u) = 
\log(T) \Pi_{\mu_1\nu_1\mu_2\nu_2}^{\epsilon}(k),
\end{equation}
where $\Pi_{\mu_1\nu_1\mu_2\nu_2}^{\epsilon}(k)$ is the residue of the
ultraviolet divergent $T=0$ contribution which is obtained
from the calculation in $n=4-2\,\epsilon$ dimensions.
The fact that both the $\log(T)$ and the ultraviolet divergent contributions
have the same structure has been also verified for the two and four-point
functions in QED \cite{BrandtFrenkelLog1} and for the two-
and three-point functions in Yang-Mills theories \cite{BrandtFrenkelLog2}.
These results are special examples of the rather general
arguments presented in \cite{BrandtFrenkelLog3}. 

From the results for the thermal one- and two graviton functions
we can write the following expression for the {\it thermal effective action} 
\begin{equation}\label{Stherm}
S_{term}[g]=\Gamma_{\mu\nu}\,h^{\mu\nu}(0)+
\int {\rm d}^4 k\,
h^{\mu_1\nu_1}(k) \Pi_{\mu_1\nu_1\,\mu_2\nu_2}(k) h^{\mu_2\nu_2}(-k) +\cdots.
\end{equation}
Here we will use Eq. (\ref{Stherm})
in order to derive the expressions for the new one- and two-graviton functions 
which arise when one uses the graviton representation
\begin{equation}
  \label{field1}
  g_{\mu\nu} \equiv \eta_{\mu\nu} + \delta g_{\mu\nu}.
\end{equation}
The corresponding expressions will be employed in the analysis
performed in the next section. 
Using Eqs. (\ref{field}) and  (\ref{field1}) one obtains the following
relation for the graviton fields in the two representations 
\begin{equation}\label{representation}
\kappa \,h_{\mu\nu}= - \delta g_{\mu\nu}
+\frac 1 2 \delta g^\alpha_\alpha \eta_{\mu\nu}
-\frac 1 2 \delta g^\alpha_\alpha \delta g_{\mu\nu} 
+\delta g_{\mu\alpha} g^\alpha_\nu
+\frac 1 8 \left(\delta g^\alpha_\alpha \right)^2 \eta_{\mu\nu}
-\frac 1 4 \delta g^{\alpha\beta} \delta g_{\beta\alpha} \eta_{\mu\nu} + \cdots.
\end{equation}
Inserting Eq. (\ref{representation}) into Eq. (\ref{Stherm}) 
and using the traceless property of $\Gamma_{\mu\nu}$ 
[cf. Eq. (\ref{tadpole})], we obtain
\begin{equation}\label{Stherm1}
S_{term}[g]=\check\Gamma_{\mu\nu}\,\delta g^{\mu\nu}(0)+
\int {\rm d}^4 k\,
\delta g^{\mu_1\nu_1}(k) \,\check \Pi_{\mu_1\nu_1\,\mu_2\nu_2}(k)\, 
\delta g^{\mu_2\nu_2}(-k) + \cdots ,
\end{equation}
where
\begin{equation}
\check\Gamma_{\mu\nu} = - \Gamma_{\mu\nu}
\end{equation}
and
\begin{equation}
\label{PiPicheck}
\begin{array}{lll}
\check\Pi ^{\mu_1 \nu_1 \,\mu_2 \nu_2} \left(k,u\right)
& =\Pi^{\mu_1 \nu_1 \,\mu_2\nu_2} \left(k,u\right)
& -\bigfrac{1}{2} \left( 
 \Pi^{\mu_1 \nu_1 \, \alpha}\,_\alpha \eta^{\mu_2\nu_2}
+\Pi^{\mu_2 \nu_2 \, \alpha}\,_\alpha \eta^{\mu_1\nu_1}
-\bigfrac{1}{2} 
 \Pi^{\alpha\;\;\beta}_{\;\;\alpha\;\;\beta}\eta^{\mu_1\nu_1}\eta^{\mu_2\nu_2}  
\right.  \\  
&  &   \;\;\;\;+\Gamma^{\mu_1 \nu_1}\eta ^{\mu_2 \nu_2}
               +\Gamma^{\mu_2 \nu_2}\eta ^{\mu_1 \nu_1}   \\  
&  & 
     \;\;\;\;-\Gamma^{\mu_1 \mu_2}\eta ^{\nu_1 \nu_2}
             -\Gamma^{\nu_1 \nu_2}\eta ^{\mu_1 \mu_2} \\
&  & 
     \biggl.
     \;\;\;\;-\Gamma^{\nu_1\mu_2}\eta ^{\mu_1 \nu_2}
             -\Gamma^{\mu_1 \nu_2}\eta ^{\nu_1 \mu_2}
      \biggr) 
\end{array}.
\end{equation}
We remark that while the derivation
of Eq. (\ref{PiPicheck}) is rather simple and general, a direct calculation
of $\check\Pi^{\mu_1 \nu_1 \,\mu_2 \nu_2} \left(k\right)$, 
on the other hand, would involve the manipulation of more complicated 
Feynman rules where the gauge fixing term from Eq. (\ref{lagrangian}) 
would contribute to all the n-graviton vertices.

\section{The graviton dispersion relations}
\label{sect4}

The thermal graviton self-energy has been employed in order
to investigate the propagation of gravitational waves in a
plasma \cite{Rebhan}. This can be done studying the poles of
the full graviton propagator (dispersion relations)
which is obtained from the effective action
\begin{equation}
     \label{effective}
S[g]=S_G[g] + S_{fix}[g] + S_{term}[g]
\end{equation}
where $S_G[g]$ is the Einstein action, $S_{fix}[g]$ is the gauge
fixing term and $S_{term}[g]$ is given by Eq. (\ref{Stherm}).
In this section we shall apply the results for the graviton self-energy up to
the sub-leading $T^2$ contributions in order to investigate the
gauge dependence of the dispersion relations.

Since the tadpole contribution to $S_{term}[g]$ yield 
a non-zero energy-momentum tensor in the Einstein equation
\begin{equation}
    \label{Einstein}
\frac{\delta S[g]}{\delta g_{\mu\nu}}=0,
\end{equation}
a self-consistent calculation of the full graviton propagator 
has to take into account a {\it curved background} \/so that
\begin{equation}\label{metric}
g_{\mu\nu}=g^{(0)}_{\mu\nu}+\delta g_{\mu\nu},
\end{equation}
where $g^{(0)}_{\mu\nu}$ is the solution of the Einstein equation
(\ref{Einstein}) and $\delta g_{\mu\nu}$ is the {\it metric fluctuation}.
From the corresponding second order variation of the effective action
\begin{equation}
  \label{second}
\delta^2 S[g] = -\frac 1 2 \int {\rm d}^4 x \sqrt{-g^{(0)}}\,
\delta g_{\mu_1\nu_1}P^{\mu_1\nu_1\,\mu_2\nu_2} \delta g_{\mu_2\nu_2},
\end{equation}
the graviton propagator can be obtained taking the inverse of 
$P^{\mu_1\nu_1\,\mu_2\nu_2}$. 

The contributions to $P^{\mu_1\nu_1\,\mu_2\nu_2}$ from
the first two terms in (\ref{effective}) are well known 
\cite{Rebhan,tHooftVeltman,BarthChristensen}. They involve 
components of the Riemann and Ricci tensors and the scalar curvature.
Restricting the analysis to a metric background which is {\it conformally flat},
the components of Riemann tensor can be expressed only in terms of
the Ricci tensor and the scalar curvature. Since Eq. (\ref{tadpole}) yields
a traceless energy-momentum tensor, the Einstein equation
(\ref{Einstein}) (with vanishing cosmological constant) implies that
the scalar curvature is zero and that the Ricci tensor is proportional
to Eq. (\ref{tadpole}). Using {\it geodesic normal coordinates} 
\/the thermal contributions to $P^{\mu_1\nu_1\,\mu_2\nu_2}$ can be
obtained from Eq. (\ref{Stherm1}) 
After a straightforward tensor algebra one obtains the following expression
\begin{equation}\label{PP}
\begin{array}{lll}
P^{\mu_1\nu_1\,\mu_2\nu_2}(k,u) & = &
\left\{\left[\left(1-\bigfrac{1}{\xi}\right)\left(
\eta^{\mu_1\nu_2} k^{\nu_1} k^{\mu_2} - \eta^{\mu_1\nu_1} k^{\mu_2} k^{\nu_2}
                                            \right)\right.\right.
\\
{}&{}&
+\bigfrac{1}{2}\left(1-\bigfrac{1}{2\xi}\right)
\eta^{\mu_1\nu_1}\eta^{\mu_2\nu_2}k^2
-\bigfrac{1}{2} \eta^{\mu_1\mu_2}\eta^{\nu_1\nu_2}k^2
\\
{}&{}&
-8\pi\, G \rho \left(\bigfrac{1}{3}
\eta^{\nu_1\mu_2} 
\left(4 u^{\mu_1} u^{\nu_2} - \eta^{\mu_1\nu_2}\right)
+\bigfrac{1}{6}
\eta^{\mu_1\nu_1}
\left(4 u^{\mu_2} u^{\nu_2} - \eta^{\mu_2\nu_2}\right)\right)
\\
{}&{}&
\Biggl.
+\left({\rm symmetrization\;\;under\;\; \mu_2 \leftrightarrow \nu_2}\right) \Biggr]
\\
{}&{}&
\biggl.
+\left({\rm symmetrization\;\;under\;\; \mu_1 \leftrightarrow \nu_1}\right)\biggr\}
- 16\pi G\check\Pi ^{\mu_1 \nu_1 \,\mu_2 \nu_2} \left(k,u\right)
\end{array},
\end{equation}
where $\check\Pi ^{\mu_1 \nu_1 \,\mu_2 \nu_2}$ is given by
Eq. (\ref{PiPicheck})  with the leading and 
the sub-leading high-temperature contributions from 
$\Pi ^{\mu_1 \nu_1 \,\mu_2 \nu_2}$ given
respectively by Eqs. (\ref{PiT4}) and (\ref{PiT2}).

Because of the coordinate invariance of the problem we have to
impose physical constraints on the metric fluctuations.
The imposition that the spin one and spin zero degrees of freedom
do not propagate constraints the metric perturbations
$\delta g_{\mu\nu}$ to be {\it transverse} \/and {\it traceless}, 
respectively\cite{GriboskyDonoghue}. 
These conditions imply that we only have to consider
the transverse and traceless components of 
$\left(P^{\alpha\beta \,\mu\nu}\right)^{-1}$
in the linear response equation
\begin{equation}\label{response}
\delta g_{\alpha\beta} = -16\pi G \left(P^{\alpha\beta \,\mu\nu}\right)^{-1}
\delta T^{\mu\nu}.
\end{equation}
An explicit basis of TT-tensors can be found imposing the TT conditions
on a general linear combination such as the one on the
right hand side of Eq. (\ref{decomposition}). It is also convenient
to choose these tensors as being idempotent and orthogonal to each other.
This leads to the following set of  TT-tensors 
\begin{equation}\label{TTtensors}
{\cal T}_I^{\mu_1\nu_1\;\mu_2\nu_2}=
\sum_{i=1}^{14}c_{I\,_{i}}T^{\mu_1\nu_1\;\mu_2\nu_2},\;\;\; I=A,B,C.
\end{equation}
where the coefficients $c_{I\,i}$ are given in the table~\ref{tab2}.

This result is in agreement with the one obtained by Rebhan in 
reference \cite{Rebhan} (except for a small misprint in the 9th line
of the 2nd row in table 2).
As a simple checkup of this result
we note that at zero temperature there is only one TT-tensor given by
\begin{equation}\label{TTT0}
{\cal T}_0^{\mu_1\nu_1\;\mu_2\nu_2}=
{\cal T}_A^{\mu_1\nu_1\;\mu_2\nu_2}+
{\cal T}_B^{\mu_1\nu_1\;\mu_2\nu_2}+
{\cal T}_C^{\mu_1\nu_1\;\mu_2\nu_2},
\end{equation}
so that in the table~\ref{tab2} the lines $1$, $4$, $8$, $12$ and $13$
(which gives the coefficients of the Lorentz covariant tensors in
the table~\ref{tab1})
must add to a Lorentz scalar (or a pure number) and all the other
lines must add to zero. Once we know a certain
set of coefficients $c_i$ in the basis given by table~\ref{tab1}
and the explicit form of the TT-tensors given by Eq. (\ref{TTtensors})
a straightforward calculation gives the following result for the 
coefficients in the basis of TT-tensors
\begin{equation}\label{TTcoeffs}
\left.
\begin{array}{lll}
c_A & = & 2\,c_1 
\\{}&{}&{}\\
c_B & = & 2\left(c_1+\bigfrac{k^2- 
\left(k\cdot u\right)^2}{k^2}\,c_2\right) 
\\{}&{}&{}\\
c_C & = & 2\left(c_1+
\bigfrac{4}{3}\bigfrac{k^2- 
\left(k\cdot u\right)^2}{k^2}\,c_2+\bigfrac{1}{3}
\bigfrac{\left(k^2- 
\left(k\cdot u\right)^2\right)^2}{\left(k^2\right)^2}\,c_3
\right)
\end{array}\right..
\end{equation}
It is interesting to note that the sub-leading contributions to the
graviton self-energy are independent of the graviton representation.
This property is satisfied because these contributions to
$\check\Pi ^{\mu_1 \nu_1 \,\mu_2 \nu_2}$ and
$\Pi ^{\mu_1 \nu_1 \,\mu_2 \nu_2}$ have the {\it same} \/TT components.
Indeed, we see from Eq. (\ref{PiPicheck}) that apart from the tadpole
contributions which have an exact $T^4$ behavior. The 
terms involving traces of $\Pi ^{\mu_1 \nu_1 \,\mu_2 \nu_2}$ are
either proportional to $\eta^{\mu_1\nu_1}$ or $\eta^{\mu_2\nu_2}$ or
both. Such terms have no components along any of the first 3 tensors
of table~\ref{tab1},  so that they give no contribution to any of the
coefficients $c_i$, $i=1,\,2,\,3$ which appear in Eqs. (\ref{TTcoeffs}). 

We have now all the basic quantities which are needed in order to express 
$P^{\mu_1\nu_1\,\mu_2\nu_2}$ in the basis of TT-tensors as follows
\begin{equation}
P^{\mu_1\nu_1\,\mu_2\nu_2}=
c_A {\cal T}_A^{\mu_1\nu_1\,\mu_2\nu_2}+
c_B {\cal T}_A^{\mu_1\nu_1\,\mu_2\nu_2}+
c_C {\cal T}_A^{\mu_1\nu_1\,\mu_2\nu_2}+
\sum_{i=4}^{14} c_i T_i^{\mu_1\nu_1\,\mu_2\nu_2}.
\end{equation}
The inverse 
\begin{equation}
\left(P^{\mu_1\nu_1\,\mu_2\nu_2}\right)^{-1}=
d_A {\cal T}_{A\;{\mu_1\nu_1\,\mu_2\nu_2}}+
d_B {\cal T}_{A\;{\mu_1\nu_1\,\mu_2\nu_2}}+
d_C {\cal T}_{A\;{\mu_1\nu_1\,\mu_2\nu_2}}+
\sum_{i=4}^{14} d_i T_{i\;{\mu_1\nu_1\,\mu_2\nu_2}}
\end{equation}
can be determined from the relation
\begin{equation}
\left(P^{\mu_1\nu_1\,\rho\sigma}\right)^{-1}\,
P^{\rho\sigma\,\mu_2\nu_2}=\frac 12\left(
\delta_{\mu_1}^{\mu_2} \delta_{\nu_1}^{\nu_2}+
\delta_{\mu_1}^{\nu_2} \delta_{\nu_1}^{\mu_2}\right).
\end{equation}
Using the transversality and idempotence of 
${\cal T}_I^{\mu\nu\,\alpha\beta}$ as well as the identities
\begin{equation}
\left.
\begin{array}{lll}
{\cal T}_{A\;\;\;\rho\sigma}^{\mu\nu}\,T_{i}^{\rho\sigma \,\alpha\beta}=0\;\;\; (i=4\cdots14) 
\\{}&{}&{}\\
{\cal T}_{B\;\;\;\rho\sigma}^{\mu\nu}\,T_{i}^{\rho\sigma \,\alpha\beta}
\left\{
\begin{array}{lll}
0 & {\rm for} & i \neq 6 \\
\neq 0 & {\rm for} & i=6
\end{array}
\right.
\\{}&{}&{}\\
{\cal T}_{C\;\;\;\rho\sigma}^{\mu\nu}\,T_{i}^{\rho\sigma \,\alpha\beta}=0
\;\;\; 
(i=4,8,10\cdots 14)
\\{}&{}&{}\\
{\cal T}_{I\;\;\;\rho\sigma}^{\mu\nu}\,T_{i}^{\rho\sigma \,\alpha\beta}=0
\;\;\; 
(i=4\cdots 14;\;I=A,B,C)
\end{array}\right. ,
\end{equation}
we obtain the following result
\begin{equation}
\left.
\begin{array}{lll}
d_A & = & \bigfrac{1}{c_A} 
\\{}&{}&{}\\
d_B & = & \bigfrac{1}{c_B}\left(
1-\bigfrac{1}{2}d_6\,c_6\,T^{\mu\nu}_{6\;\;\rho\sigma} 
T_6^{\rho\sigma\alpha\beta} {\cal T}_{B\,\mu\nu\alpha\beta}\right)
\\{}&{}&{}\\
d_C & = & \bigfrac{1}{c_C}\left(
1-\sum_{(i,j=5,6,7,9)} d_i\,c_j\,T^{\mu\nu}_{i\;\;\rho\sigma} 
T_j^{\rho\sigma\alpha\beta} {\cal T}_{C\,\mu\nu\alpha\beta}\right)
\end{array}\right..
\end{equation}

We can now investigate the poles of the TT components of the
propagator from the solution of the equations $c_I=0,\;I=A,B,C$.
Using the Eqs. (\ref{TTcoeffs}) with $c_1$, $c_2$ and $c_3$
determined from the decomposition of (\ref{PP}) in the basis
of table~\ref{tab1}, the equations associated with the modes
$A$, $B$ and $C$ can be written respectively in the form
\begin{equation}
  \label{modes} 
\left.
\begin{array}[h]{lcl}
{k^2}
& = & \bigfrac{16\pi\, G\rho}{
1-\displaystyle{{\bigfrac{64\pi}{15} \, {G\, T^2}}}
\left({1- \xi}\right)}
\left[
\left(
\displaystyle{\bigfrac 5 9} 
+ \displaystyle{\bigfrac 1 2}\,{r}^{4} L-\displaystyle{\bigfrac 1 6}\,{r}^{2}
\right)+
\bigfrac{8\, k^2}{\pi^2 T^2}\left(8\,{r}^{2}\,L + \bigfrac{1}{2}{r}^4 L -
\bigfrac{1}{6}\,{r}^{2}-4\right)
\right]
 \\
{}&{}&\\
{k^2}
& = & \bigfrac{16\pi\, G\rho}{
1+\displaystyle{{\bigfrac{64\pi}{15} \, {G\, T^2}}}
\left({1- \xi}\right)}
\left[
\left(\displaystyle{\bigfrac 2 9} - 2\,{r}^{4} L 
+ \displaystyle{\bigfrac 2 3}\,{r}^{2} + {\bigfrac {10}{9}}\,
\bigfrac{1}{{r}^{2}}\right)+\bigfrac{8\, k^2}{\pi^2 T^2}\left(
-5\,{r}^{2}L\,-2\,L\,{r}^{4}+\bigfrac{2}{3}\,{r}^{2}\right)
\right]
 \\
{}&{}&\\
{k^2}
& = & \bigfrac{16\pi\, G\rho}{
1-\displaystyle{{\bigfrac{64\pi}{15} \, {G\, T^2}}}
\left({1- \xi}\right)}
\left[\left(
{\displaystyle{\bigfrac {8}{9}}}
+3\,{r}^{4} L -{r}^{2} 
+ \displaystyle{{\bigfrac {28}{27}}\,\bigfrac{1}{{r}^{2}}}\right)
+\bigfrac{8\, k^2}{\pi^2 T^2}\left(1-6\,{r}^{2}L\, + 3\,L\,{r}^{4} - {r}^{2}\right)\right]
\end{array} \right. ,
\end{equation}
where
\begin{equation}\label{rL}
r^2 \equiv \bigfrac{k^2}{\teb{k}\cdot\teb{k}};\;\;\;
L \equiv \bigfrac{k_0}{2|\teb{k}|} \log{\bigfrac{k_0+|\teb{k}|}{k_0-|\teb{k}|}}-1\;\;
\end{equation}
The Eqs. (\ref{modes}) reduces to the Eqs. (6.2) of reference
\cite{Rebhan} in the special case when
the sub-leading terms proportional to $G\,T^2$ or 
$k^2/T^2$ are neglected. 

In view  of the constraints imposed by the important condition
(\ref{X0PiT2}), the gauge dependent denominators in Eqs. (\ref{modes})
have a very simple structure. Since we assume that $G\, T^2\ll 1$,
the momentum-independent denominators can be expanded perturbatively.
We thus see that all the gauge dependent $T^2$
sub-leading contributions to the dispersion relations give effectively
corrections of order $G^2$, which are of the same  order as the two-loop
contributions which we have disregarded.
Hence, we conclude that to one-loop order, the dispersion relations
are effectively gauge independent.

The solution of the one-loop dispersion
relations can therefore be obtained from Eqs. (\ref{modes}), by setting
the denominators equal to one.
These solutions have been obtained in the reference 
\cite{Rebhan} in the leading high temperature approximation. 
In order to illustrate the magnitude of the $T^2$ sub-leading contributions
let us consider the solution of Eqs. (\ref{modes}) for real values of
$k_0$ and $\teb{k}$. This corresponds to the propagation of waves supported by
the graviton plasma. In the figure 3, where
$\bar\omega \equiv k_0/    \left(16\pi G\rho\right)^{1/2}$
and $\bar k\equiv |\teb{k}|/    \left(16\pi G\rho\right)^{1/2}$,
we show the numerical solutions corresponding to the modes 
$A$, $B$ and $C$ and compare the leading results 
with the contributions which include the sub-leading $T^2$
corrections.
We can see from these diagrams that for all TT-modes, the
dispersion curves begin at a common plasma
frequency $\bar \omega_{pl}$ and become asymptotically
parallel to the light cone.

The behavior of the dispersion relation can be determined analytically
in the limiting cases of very small and very large momenta.
When $\bar k\rightarrow 0$, the common form of the dispersion 
relations is given by
\begin{equation}\label{plasmaF}
\bar\omega_{pl}   =   
\sqrt{\frac{22}{45}}\left(1 -  \frac{224}{75} \pi G\, T^2 \right)
\end{equation}
For large momenta such that $\bar k \gg 1$, the asymptotic forms of the dispersion
relations become respectively
\begin{equation}
\begin{array}{lll}
\bar\omega_A & = & \bar k_A + \bigfrac{5}{18\bar k_A}\left(1-\frac{256}{15} \pi G\, T^2\right) \\
\bar\omega_B & =   & \bar k_B + \sqrt{\bigfrac{5}{18}}  \\
\bar\omega_C & =   & \bar k_C + \sqrt{\bigfrac{7}{27}}\left(1+\bigfrac{32}{15}\pi G\, T^2\right) \\
\end{array}.
\end{equation}
The small wavelength limit given by the above relations
can be understood by noticing that in this case one probes the plasma
at small distances, where the medium effects on the free dispersion relations are 
relatively unimportant.
On the other hand, for long wavelengths Eq. (\ref{plasmaF}) gives 
a substantial modification of the free dispersion relation
due to the collective phenomena in the plasma.
However, this modification is sensitive to the curvature of space which we have
neglected in our analysis. (Such an effect yields corrections of magnitude
$(G\,T^4)(G\,T^2)$, which are formally of the same order as the two-loop contributions).
This is an interesting issue which deserves further study.

\acknowledgements{We would like to thank ${\rm CNP_Q}$, Brasil, for a grant
and Prof. J. C. Taylor for a helpful correspondence}.

\newpage
\appendix

\section{}
\label{appA}

Here we show explicitly how to extend the method of Barton 
amplitudes in order to account for the contributions which 
arises from the quadratic denominators in the general gauge free propagator. 
We illustrate this technique by considering the following integral
\begin{equation}\label{genint}
I=\int{\rm d}^3\teb{q} \int_{-i\infty+\delta}^{+i\infty+\delta}
\frac{{\rm d} z}{2\pi i} N(z)
\left[\frac{t(q; p)}{\left(q\cdot q\right)^{i}
                               \left(p\cdot p\right)^{j}}
      + (z\leftrightarrow -z) \right],
\end{equation}
where $p\equiv q+k,\; k_0=2 m \pi i, m=0,\pm 1,\pm 2\cdots$, 
$q=(z,\teb{q})$ and \hbox{${i},\,{j}=1,2$}.
This is the most general kind of integral which contributes
to the two-point function when the imaginary time formalism
is employed. The generalization to higher point functions is
straightforward. The numerator $t(q; p)$ comes from the 
graviton vertices and from the numerator of the free propagator.

Since the integration in (\ref{genint}) is over all values of $\teb{q}$, 
it is more convenient to make the change of variables 
$\teb{q}\leftrightarrow -\teb{q}$ in all the terms$(z\leftrightarrow -z)$
so that 
\begin{equation}\label{genint2}
I=\int{\rm d}^3\teb{q} \int_{-i\infty+\delta}^{+i\infty+\delta}
\frac{{\rm d} z}{2\pi\,i} N(z)
\left[\frac{t(q; p)}{\left(q\cdot q\right)^{i}
                               \left(p\cdot p\right)^{j}}
      + q\leftrightarrow -q \right].
\end{equation}
Factorizing the denominators in (\ref{genint2}) we can write
\begin{equation}\label{genint3}
\begin{array}{ll}
I=\bigint{\rm d}^3\teb{q} \bigint_{-i\infty+\delta}^{+i\infty+\delta}
\bigfrac{{\rm d} z}{2\pi\,i} N(z) & \left[
\bigfrac{1}{\left(z+|\teb{q}|\right)^{i}}
\bigfrac{1}{\left(z+k_0+|\teb{p}|\right)^{j}} 
\right. 
\\ {}&{} \\
& \left.
\bigfrac{t(q; p)}{\left(z-|\teb{q}|\right)^{i}
\left(z+k_0-|\teb{p}|\right)^{j}}
+ q\leftrightarrow -q \right]
\end{array}.
\end{equation}
The $z$ integration is now readily performed using the Cauchy theorem
and closing the contour in the right hand side plane where
the only poles are located at $z=|\teb{q}|$ and $z=|\teb{p}|-k_0$
($k_0$ is a pure imaginary quantity at this stage of the calculation).
In this way we obtain
\begin{equation}\label{genint4}
\begin{array}{ll}
I=-\bigint {\rm d}^3\teb{q} & 
\left\{\bigfrac{1}{({i}-1)!}
\lim_{q_0\rightarrow|\scriptsize{\teb{q}}|}\bigfrac{\partial^{{i}-1}}{
\partial q_0^{{i}-1}}\left(\bigfrac{N(q_0)}{(q_0+|\teb{q}|)^{i}}
                   \bigfrac{t(q;p)}{(p\cdot p)^{j}}\right) 
\right.
\\{}&{}\\
{}&\left. 
+ \bigfrac{1}{({j}-1)!}
\lim_{q_0\rightarrow|\scriptsize{\teb{p}}|-k_0}\bigfrac{\partial^{{j}-1}}{
\partial q_0^{{j}-1}}\left(\bigfrac{N(q_0)}{(q_0+k_0+|\teb{p}|)^{j}}
                   \bigfrac{t(q;p)}{(q\cdot
                     q)^{i}}\right)
+ q\leftrightarrow -q \right\}
\end{array}.
\end{equation}
Performing the change of variables $\teb{q}\rightarrow\teb{q} - \teb{k}$ in the
second term of (\ref{genint4}) we can write
\begin{equation}\label{genint5}
\begin{array}{ll}
I=-\bigint {\rm d}^3\teb{q} & 
\left\{\bigfrac{1}{({i}-1)!}
\lim_{q_0\rightarrow|\scriptsize{\teb{q}}|}\bigfrac{\partial^{{i}-1}}{
\partial q_0^{{i}-1}}\left(\bigfrac{N(q_0)}{(q_0+|\teb{q}|)^{i}}
                   \bigfrac{t(q;p)}{(p\cdot p)^{j}}\right) 
\right.
\\{}&{}\\
{}&\left. 
+ \bigfrac{1}{({j}-1)!}
\lim_{q_0\rightarrow|\scriptsize{\teb{q}}|-k_0}\bigfrac{\partial^{{j}-1}}{
\partial q_0^{{j}-1}}\left(\bigfrac{N(q_0)}{(q_0+k_0+|\teb{q}|)^{j}}
                   \bigfrac{t(q_0,\teb{q}-\teb{k};q_0+k_0,\teb{q})}
                           {(q_0^2-|\teb{q}-\teb{k}|^2)^{i}}\right)
+ q\leftrightarrow -q \right\}
\end{array}.
\end{equation}
Finally, using the property $N(q_0+k_0)=N(q_0)$ and the
symmetry $q\leftrightarrow -q$ we obtain
\begin{equation}\label{genint6}
\begin{array}{ll}
I=-\bigint {\rm d}^3\teb{q} & 
\left\{\bigfrac{1}{({i}-1)!}
\bigfrac{\partial^{{i}-1}}{
\partial q_0^{{i}-1}}\left(\bigfrac{N(q_0)}{(q_0+|\teb{q}|)^{i}}
                   \bigfrac{t(q;p)}{(p\cdot p)^{j}}\right) 
\right.
\\{}&{}\\
{}&\left. 
+ \bigfrac{1}{({j}-1)!}
\bigfrac{\partial^{{j}-1}}{
\partial q_0^{{j}-1}}\left(\bigfrac{N(q_0)}{(q_0+|\teb{q}|)^{j}}
                           \bigfrac{t(-p;q)}
                           {(p\cdot p)^{i}}\right)
+ q\leftrightarrow -q \right\}_{q\cdot q = 0}
\end{array}.
\end{equation}
The special case when ${i}={j}=1$ gives the known result
\begin{equation}\label{genint7}
I=-\int \frac{{\rm d}^3\teb{q}}{2\,|\teb{q}|}N(|\teb{q}|) 
\left[\frac{t(q;p)+t(-p;q)}{(p\cdot p)}
      + q\leftrightarrow -q \right]_{q\cdot q = 0},
\end{equation}
where the expression inside the bracket is a typical contribution 
to the on-shell forward scattering amplitude. Although the derivatives
in the general expression (\ref{genint6}) makes it much more 
difficult to be handled, it is straightforward to deal with such kind
of expressions using a {\it computer algebra program}. 

\section{}
\label{appB}

The Feynman rules are obtained inserting
Eq. (\ref{field}) and the corresponding perturbative expansion
of the inverse
\begin{equation}
\left(\sqrt{-g}\,g^{{\mu}{\nu}}\right)^{-1}
                   =\eta_{{\mu}{\nu}}-\kappa\, h_{{\mu}{\nu}}
                   +\kappa^2\, h_{{\mu}{\alpha}}\, h_{{\alpha}{\nu}}
                   -\kappa^3\, h_{{\mu}{\alpha}}\, h_{{\alpha}{\beta}}
                    \, h_{{\beta}{\nu}}
                   +O(\kappa^4)
\end{equation}
into Eq. (\ref{lagrangian}). The contributions of order 0 in $\kappa$
yields the graviton propagator given by Eq. (\ref{propag}).

The third term in Eq. (\ref{lagrangian}) yields the following
expressions for the ghost propagator and the graviton-ghost-ghost
vertex
\begin{equation}
{\cal D}^{ghost}_{{\mu}{\nu}}(k)=\frac{\eta_{{\mu}{\nu}}}{k^2},
\end{equation}
\begin{equation}
{\cal V}^{ghost}_{{\mu_1}{\nu_1}\,{\mu_2}{\mu_3}}
(k_1,k_2,k_3)=
\frac\kappa 2
\left[\eta_{{\mu_2}{\mu_3}}({k_2}_{\mu_1} {k_3}_{\nu_1}
                           +{k_3}_{\mu_1} {k_2}_{\nu_1})-
       (\eta_{{\mu_1}{\mu_2}}{k_1}_{\nu_1} +
        \eta_{{\nu_1}{\mu_2}}{k_1}_{\mu_1} ){k_2}_{\mu_3}\right].
\end{equation}

All the graviton self-couplings are generated only from the
first term in Eq. (\ref{lagrangian}). The corresponding Feynman
rules for the three- and  four-graviton couplings 
are given respectively by the following expressions
 
\begin{equation}\label{3grav}
\begin{array}{lll}
{}&{}&
{\cal V}^3_{{\mu_1}{\nu_1}\,{\mu_2}{\nu_2}\,{\mu_3}{\nu_3}}(k_1,k_2,k_3)=\\
{}&{}&
\bigfrac \kappa 4 \left[
-4\,{k_2}_{{\mu_3}}\,{k_3}_{{\nu_2}}\,\eta_{{\mu_1}{\mu_2}}\,\eta_{{\nu_1}{\nu_3}}
-{k_2}\cdot{k_3}\,\eta_{{\mu_1}{\mu_3}}\,\eta_{{\nu_1}{\nu_3}}\,\eta_{{\mu_2}{\nu_2}}
+2\,{k_2}\cdot {k_3}\,\eta_{{\mu_1}{\nu_2}}\,\eta_{{\nu_1}{\nu_3}}\,\eta_{{\mu_2}{\mu_3}} 
\right. \\
{}&{}&
+2\,{k_2} \cdot {k_3}\,\eta_{{\mu_1}{\mu_2}}\,\eta_{{\nu_1}{\mu_3}}\,\eta_{{\nu_2}{\nu_3}} 
-2\,{k_2}_{{\mu_1}}\,{k_3}_{{\nu_1}}\,\eta_{{\mu_2}{\mu_3}}\,\eta_{{\nu_2}{\nu_3}}
\\
{}&{}&
-{k_2}\cdot{k_3}\,\eta_{{\mu_1}{\mu_2}}\,\eta_{{\nu_1}{\nu_2}}\,\eta_{{\mu_3}{\nu_3}}
+{k_2}_{{\mu_1}}\,{k_3}_{{\nu_1}}\,\eta_{{\mu_2}{\nu_2}}\,\eta_{{\mu_3}{\nu_3}} \\
{}&{}&
+({\rm symmetrization}\;\;{\rm under}\;\;({\mu_1}\leftrightarrow{\nu_1}),\;\;
                        ({\mu_2}\leftrightarrow{\nu_2}),\;\;
                        ({\mu_3}\leftrightarrow{\nu_3}))\;\;\left.\right]\\
{}&{}&
+{\rm permutations}\;\;{\rm of}\;\;
(k_1,{\mu_1},{\nu_1}),\;\;(k_2,{\mu_2},{\nu_2}),\;\;(k_3,{\mu_3},{\nu_3})\;\;,
\end{array}
\end{equation}

\begin{equation}
\begin{array}{lll}
{}&{}&
{\cal V}^4_{{\mu_1}{\nu_1}\,{\mu_2}{\nu_2}\,{\mu_3}{\nu_3}\,{\mu_4}{\mu_4}}
(k_1,k_2,k_3,k_4) = \\
{}&{}&
\bigfrac{\kappa^2}{4}
\left[
-2\,{k_3}_{{\mu_2}}\,{k_4}_{{\nu_2}}\,\eta_{{\mu_1}  {\nu_3}}\,
\eta_{{\nu_1}  {\mu_4}}\,\eta_{{\mu_4}  {\mu_3}}
+ {k_3}_{{\mu_2}}\,{k_4}_{{\nu_2}}\,\eta_{{\mu_1}  {\mu_3}}\,
   \eta_{{\nu_1}  {\nu_3}}\,\eta_{{\mu_4}  {\mu_4}} \right.
\\ {}&{}&
-{k_3} \cdot {k_4}\,\eta_{{\mu_1}  {\mu_3}}\,\eta_{{\nu_1}  {\nu_2}}\,
   \eta_{{\mu_4}  {\mu_4}}\,\eta_{{\mu_2}  {\nu_3}}
-4\,{k_3}_{{\mu_4}}\,{k_4}_{{\nu_3}}\,\eta_{{\mu_1}  {\mu_3}}\,
   \eta_{{\nu_1}  {\nu_2}}\,\eta_{{\mu_2}  {\mu_4}}
\\ {}&{}&
+2\,{k_3} \cdot {k_4}\,\eta_{{\mu_1}  {\nu_3}}\,\eta_{{\nu_1}  {\nu_2}}\,
   \eta_{{\mu_4}  {\mu_3}}\,\eta_{{\mu_2}  {\mu_4}}
- {k_3} \cdot {k_4}\,\eta_{{\mu_1}  {\mu_3}}\,\eta_{{\nu_1}  {\nu_3}}\,
   \eta_{{\mu_4}  {\mu_2}}\,\eta_{{\nu_2}  {\mu_4}}
\\ {}&{}&
+ 2\,{k_3} \cdot {k_4}\,\eta_{{\mu_1}  {\mu_3}}\,\eta_{{\nu_1}  {\mu_4}}\,
   \eta_{{\mu_2}  {\nu_3}}\,\eta_{{\nu_2}  {\mu_4}}
+   {k_3}_{{\mu_2}}\,{k_4}_{{\nu_2}}\,\eta_{{\mu_1}  {\mu_4}}\,\eta_{{\nu_1}  {\mu_4}}\,
   \eta_{{\mu_3}  {\nu_3}}
\\ {}&{}&
- {k_3} \cdot {k_4}\,\eta_{{\mu_1}  {\mu_4}}\,
   \eta_{{\nu_1}  {\nu_2}}\,\eta_{{\mu_2}  {\mu_4}}\,\eta_{{\mu_3}  {\nu_3}}
- 2\,{k_3}_{{\mu_2}}\,{k_4}_{{\nu_2}}\,\eta_{{\mu_1}  {\mu_3}}\,
   \eta_{{\nu_1}  {\mu_4}}\,\eta_{{\nu_3}  {\mu_4}}
\\ {}&{}&
+ 2\,{k_3} \cdot {k_4}\,\eta_{{\mu_1}  {\mu_3}}\,\eta_{{\nu_1}  {\nu_2}}\,
   \eta_{{\mu_4}  {\mu_2}}\,\eta_{{\nu_3}  {\mu_4}}
\\ {}&{}&
+({\rm symmetrization}\;\;{\rm under}\;\;({\mu_1}\leftrightarrow{\nu_1}),\;\;
                               ({\mu_2}\leftrightarrow{\nu_2}),\;\;
                               ({\mu_3}\leftrightarrow{\nu_3}),\;\;
                               ({\mu_4}\leftrightarrow{\mu_4}))\;\;\left.\right]
\\ {}&{}&
+{\rm permutations}\;\;
{\rm of}\;\;(k_1,{\mu_1},{\nu_1}),\;\;(k_2,{\mu_2},{\nu_2}),\;\;(k_3,{\mu_3},
{\nu_3}),\;\;
(k_4,{\mu_4},{\mu_4})\;\;.
\end{array}
\end{equation}
As usual, we have energy-momentum conservation at the vertices, where all
momenta are defined to be inwards.

\section{Gravitational {\rm \'{}t} Hooft identities}
\label{appC}
The imaginary time formalism at finite temperature follows closely the corresponding
formalism at $T=0$. Consequently, the \'{}t Hooft identities at finite $T$ would be similar
to the ones at $T=0$, were it not for the presence of 1-particle tadpole contributions
(such terms vanish at $T=0$ in the dimensional regularization scheme). However, since
the tadpole terms are proportional to $T^4$, they do not affect the identities involving
the sub-leading contributions. To derive these, we start from the action
\begin{equation}
  \label{Iaction}
I =  \int {\rm d}^4 x  {\rm d}^4 y h_{\mu\nu}(x) S_{sub}^{\mu\nu\,\alpha\beta}(x-y)
                              h_{\alpha\beta}(y) +
  \int {\rm d}^4 x  {\rm d}^4 y J^{\mu\nu}(x) X_{\mu\nu\, \lambda}(x-y)
                              \eta^{\lambda}(y) + \cdots.
\end{equation}
Here $S_{sub}^{\mu\nu\,\alpha\beta}$ denotes the sub-leading contributions
to the graviton 2-point function and $X_{\mu\nu\, \lambda}$ represents
the tensor generated by a gauge transformation of the graviton field which
is given 
to lowest order by Eq. (\ref{X0}) in the momentum space. $J^{\mu\nu}$ is an
external source, $\eta^{\lambda}$ represents the ghost field and $\cdots$ stand
for terms which are not relevant for our purpose. The \'{}t Hooft identity
involving the graviton self-energy function is a consequence of the BRST
invariance of the action $I$:
\begin{equation}
\label{brst}
\int{\rm d}^4 x\frac{\delta I}{\delta J^{\mu\nu}(x)} 
              \frac{\delta I}{\delta h_{\mu\nu}(x)} = 0
\end{equation}
To lowest order, Eq. (\ref{brst}) is equivalent to the relation Eq. 
(\ref{X0S0}). In general, Eq. (\ref{brst}) implies the generalized \'{}t Hooft identity
\begin{equation}
\label{gentHooft1}
X_{\mu\nu\; \lambda}\, S_{sub}^{\mu\nu\,\alpha\beta} = 0 ,
\end{equation}
which can be written to second order as
\begin{equation}
\label{gentHooft2}
X^{(0)}_{\mu\nu\,\lambda}\, \Pi_{sub}^{\mu\nu\,\alpha\beta} = 
- X^{(1)}_{\mu\nu\,\lambda}\, S^{(0)\; \mu\nu\,\alpha\beta}.
\end{equation}
Using Eq. (\ref{X0S0}), we see that (\ref{gentHooft2}) leads immediately to the
\'{}t Hooft identity (\ref{tHooft}).

In order to derive Eq. (\ref{X0PiT2}), we shall need to evaluate the tensor 
$X^{(1)}_{\mu\nu\, \lambda}$ which appears in Eq. (\ref{gentHooft2}). This tensor
may be represented by the diagram shown in Fig. 4, where the ghost-graviton-source
vertex is given in the Appendix A of ref. \cite{Delbourgo}.
Using the forward scattering amplitude method, we obtain
the following structure for the $T^2$ contributions to
$X^{(1)}_{\mu\nu\, \lambda}$
\begin{equation}
X^{(1)\; T^2}_{\mu\nu\, \lambda} =  X_{\mu\nu}^{(0)\gamma} \;
\kappa^2
\bigfrac{T^2}{48 \pi}\bigfrac{5-\xi}{4}\;
\bigint{\rm d}\Omega
\left(2 \eta_{\gamma\lambda} + 
\bigfrac{k_\gamma Q_\lambda + Q_\gamma k_\lambda}{k\cdot Q} - 
\bigfrac{k^2 Q_\gamma Q_\lambda}{(k\cdot Q)^2}
\right)
\end{equation}
Using this form and Eq. (\ref{X0S0}), it is clear
that $X^{(1)\; T^2}_{\mu\nu\;\lambda}$ is orthogonal to 
$S^{(0)\;\mu\nu\,\alpha\beta}$, so that the relation
(\ref{X0PiT2}) follows at once from the identity (\ref{gentHooft2}).

\begin{table}[t]
\begin{center}
$$
  \begin{array}{l}
T^{\mu_1\nu_1 \; \mu_2\nu_2}_1=\eta ^{\mu_2 \nu_1}\,\eta ^{\nu_2 \mu_1}+
\eta ^{\mu_2 \mu_1}\,\eta ^{\nu_2 \nu_1}   \\ 
T^{\mu_1\nu_1 \; \mu_2\nu_2}_2=u^{\mu_1} \,\left( u^{\nu_2}
\,\eta ^{\mu_2 \nu_1}+u^{\mu_2} \,\eta ^{\nu_2 \nu_1}\right) +
u^{\nu_1} \,\left( u^{\nu_2} \,\eta ^{\mu_2 \mu_1}+u^{\mu_2} \,\eta ^{\nu_2 \mu_1}
\right)   \\ 
T^{\mu_1\nu_1 \; \mu_2\nu_2}_3=u^{\mu_2} \,u^{\nu_2} \,u^{\mu_1} \,u^{\nu_1}   \\ 
T^{\mu_1\nu_1 \; \mu_2\nu_2}_4=\eta ^{\mu_2 \nu_2}\,\eta ^{\mu_1 \nu_1}   \\ 
T^{\mu_1\nu_1 \; \mu_2\nu_2}_5=u^{\mu_1} \,u^{\nu_1} \,\eta ^{\mu_2 \nu_2}+
u^{\mu_2} \,u^{\nu_2} \,\eta ^{\mu_1 \nu_1}  \\ 
T^{\mu_1\nu_1 \; \mu_2\nu_2}_6=u^{\nu_2} \,\left( k^{\nu_1} \,
\eta ^{\mu_2 \mu_1}+k^{\mu_1} \,\eta ^{\mu_2 \nu_1}\right) +
k^{\nu_2} \,\left( u^{\nu_1} \,\eta ^{\mu_2 \mu_1}+u^{\mu_1} \,\eta ^{\mu_2 \nu_1}
\right)  \\
\;\;\;\;\;\;\;\;\;\;\;\;\;\,\;\,\;\;\;
+\;u^{\mu_2} \,\left( k^{\nu_1} \,\eta ^{\nu_2 \mu_1}+
k^{\mu_1} \,\eta ^{\nu_2 \nu_1}\right) +k^{\mu_2} \,\left( u^{\nu_1}
\,\eta ^{\nu_2 \mu_1}+u^{\mu_1} \,\eta ^{\nu_2 \nu_1}\right)   \\ 
T^{\mu_1\nu_1 \; \mu_2\nu_2}_7=k^{\nu_1} \,u^{\mu_2} \,u^{\nu_2} \,u^{\mu_1} +
k^{\mu_1} \,u^{\mu_2} \,u^{\nu_2} \,u^{\nu_1} +k^{\nu_2} \,u^{\mu_2} \,u^{\mu_1} \,u^{\nu_1} +
k^{\mu_2} \,u^{\nu_2} \,u^{\mu_1} \,u^{\nu_1}   \\ 
T^{\mu_1\nu_1 \; \mu_2\nu_2}_8=k^{\nu_2} \,k^{\nu_1} \,\eta ^{\mu_2 \mu_1}+
k^{\nu_2} \,k^{\mu_1} \,\eta ^{\mu_2 \nu_1}+
k^{\mu_2} \,k^{\nu_1} \,\eta ^{\nu_2 \mu_1}+
k^{\mu_2} \,k^{\mu_1} \,\eta ^{\nu_2 \nu_1}  \\ 
T^{\mu_1\nu_1 \; \mu_2\nu_2}_9=k^{\mu_1} \,k^{\nu_1} \,u^{\mu_2} \,u^{\nu_2} +
k^{\mu_2} \,k^{\nu_2} \,u^{\mu_1} \,u^{\nu_1}   \\ 
T^{\mu_1\nu_1 \; \mu_2\nu_2}_{10}=\left( k^{\nu_2} \,u^{\mu_2} +
k^{\mu_2} \,u^{\nu_2} \right) \,\left( k^{\nu_1} \,u^{\mu_1} +k^{\mu_1} \,u^{\nu_1} \right)   \\ 
T^{\mu_1\nu_1 \; \mu_2\nu_2}_{11}=k^{\nu_2} \,k^{\mu_1} \,k^{\nu_1} \,u^{\mu_2} +
k^{\mu_2} \,k^{\mu_1} \,k^{\nu_1} \,u^{\nu_2} +k^{\mu_2} \,k^{\nu_2} \,k^{\nu_1} \,u^{\mu_1} + 
k^{\mu_2} \,k^{\nu_2} \,k^{\mu_1} \,u^{\nu_1}   \\ 
T^{\mu_1\nu_1 \; \mu_2\nu_2}_{12}=k^{\mu_2} \,k^{\nu_2} \,k^{\mu_1} \,k^{\nu_1}   \\ 
T^{\mu_1\nu_1 \; \mu_2\nu_2}_{13}=k^{\mu_1} \,k^{\nu_1} \,\eta ^{\mu_2 \nu_2}+
k^{\mu_2} \,k^{\nu_2} \,\eta ^{\mu_1 \nu_1}  \\ 
T^{\mu_1\nu_1 \; \mu_2\nu_2}_{14}=\left( k^{\nu_1} \,u^{\mu_1} +k^{\mu_1} \,u^{\nu_1}
\right) \,\eta^{\mu_2 \nu_2}+\left( k^{\nu_2} \,u^{\mu_2} +
k^{\mu_2} \,u^{\nu_2} \right) \,\eta^{\mu_1 \nu_1} \\
   \end{array}
$$
\smallskip
\caption{A basis of 14 independent tensors\label{tab1}}
\end{center}
\end{table}

\begin{table}[htbp]
    \begin{tabular}{|c|c|c|c|}
$i$&$c_{A\,_{i}}$&$c_{B\,_{i}}$&$c_{C\,_{i}}$ \\ \hline 
$1$&$\bigfrac{1}{2}$&$0$&$0$ \\ 
$2$&$\bigfrac{1}{2}\,{\bigfrac {{k^2}}{(k\cdot u)^{2}-{k^2}}}$& 
$-\bigfrac{1}{2}\,{\bigfrac {{k^2}}
{(k\cdot u)^{2}-{k^2}}}$&$0$ \\ 
$3$&$\bigfrac{1}{2}\,
{\bigfrac{{{(k^2)}}^{2}}
{\left ((k\cdot u)^{2}-{k^2}\right )^{2}}}$& 
$-2\,{\bigfrac 
{{(k^2)}^{2}}{\left ((k\cdot u)^{2}-{k^2}\right )^{2}}}$&$\bigfrac{3}{2}\,{
\bigfrac {(k^2)^2}{\left ((k\cdot u)^{2}-{k^2}\right )^{2}}}
$ \\ 
$4$&$-\bigfrac{1}{2}$&$0$&$\bigfrac{1}{6}$ \\ 
$5$&$-\bigfrac{1}{2}\,
{\bigfrac {{k^2}}{(k\cdot u)^{2}-{k^2}}}$&$0$& 
$\bigfrac{1}{2}\,{\bigfrac {{k^2}}
{{{(k\cdot u)}}^{2}-{k^2}}}$ \\ 
$6$&$-\bigfrac{1}{2}\,{\bigfrac {{k\cdot u}}
{{{(k\cdot u)}}^{2}-{k^2}}}$&$\bigfrac{1}{2}\,
{\bigfrac{{k\cdot u}}{(k\cdot u)^{2}-
{k^2}}}$&$0$ \\ 
$7$&$-\bigfrac{1}{2}\,{\bigfrac {{k^2}\,{k\cdot u}}
{\left ((k\cdot u)^{2}-{k^2}\right )^{2}}}$&$2\,{\bigfrac {{k^2}\,
{k\cdot u}}{\left ((k\cdot u)^{2}-
{k^2}\right )^{2}}}$&$-\bigfrac{3}{2}\,{\bigfrac 
{{k^2}\,{k\cdot u}}{\left ((k\cdot u)^{2}-{k^2}\right )^{2}}}
$ \\ 
$8$&$\bigfrac{1}{2}\,\left ((k\cdot u)^{2}-{k^2}\right )^{-1
}$&$-\bigfrac{1}{2}\,{\bigfrac {(k\cdot u)^{2}}{{k^2}\,\left ((k\cdot u)^{2}-
{k^2}\right )}}$&$0$ \\ 
$9$&$\bigfrac{1}{2}\,{\bigfrac {2\,(k\cdot u)^{2}-
{k^2}}{\left ((k\cdot u)^{2}-{k^2}\right )^{2}}}$& 
$-2\,{\bigfrac {{{(k\cdot u)}}^{2}}
{\left ((k\cdot u)^{2}-{k^2}\right )^{2}}}$& 
$\bigfrac{1}{2}\,{\bigfrac {2\,(k\cdot u)^{2}+
{k^2}}{\left ((k\cdot u)^{2}-{k^2}
\right )^{2}}}$ \\ 
$10$&$\bigfrac{1}{2}\,{\bigfrac {{k^2}}
{\left({{(k\cdot u)}}^{2}-{k^2}\right )^{2}}}$& 
$-\bigfrac{1}{2}\,{\bigfrac {3\,(k\cdot u)^{2}+
{k^2}}{\left ((k\cdot u)^{2}-{k^2}\right )^{2}}}$& 
$\bigfrac{3}{2}\,{\bigfrac{{{(k\cdot u)}}^{2}}
{\left ((k\cdot u)^{2}-{k^2}\right )^{2}}}
$ \\ 
$11$&$-\bigfrac{1}{2}\,{\bigfrac {{k\cdot u}}{\left ((k\cdot u)^{2}-
{k^2}\right )^{2}}}$& 
${\bigfrac {\left ((k\cdot u)^{2}+{k^2}\right)
{k\cdot u}}{{k^2}\,\left ((k\cdot u)^{2}-{k^2}\right )^{2}}}$&$-
\bigfrac{1}{2}\,{\bigfrac {\left (2\,(k\cdot u)^{2}+{k^2}\right ){k\cdot u}}
{{k^2}\,\left ((k\cdot u)^{2}-{k^2}\right )^{2}}}
$ \\ 
$12$&$\bigfrac{1}{2}\,\left ((k\cdot u)^{2}-{k^2}\right )^{-
2}$&$-2\,{\bigfrac {(k\cdot u)^{2}}{{k^2}\,\left ((k\cdot u)^{2}-
{k^2}\right )^{2}}}$& 
$\bigfrac{1}{6}\,{\bigfrac {4\,{{(k\cdot u)}}^{4}+4\,(k\cdot u)^{2}
{k^2}+(k^2)^2}{(k^2)^2\left ((k\cdot u)^{2}-
{k^2}\right )^{2}}}$ \\ 
$13$&$-\bigfrac{1}{2}\,\left ((k\cdot u)^{2}-
{k^2}\right )^{-1}$&$0$&$\bigfrac{1}{6}\,{\bigfrac {2\,(k\cdot u)^{2}+{k^2}}
{{k^2}\,\left ((k\cdot u)^{2}-{k^2}\right )}}
$ \\ 
$14$&$\bigfrac{1}{2}\,{\bigfrac {{k\cdot u}}{(k\cdot u)^{2}-{k^2}}}$& 
$0$&$-\bigfrac{1}{2}\,{\bigfrac {{k\cdot u}}{(k\cdot u)^{2}-{k^2}}}$
    \end{tabular}
    \caption{Components of the transverse traceless tensors}
    \label{tab2}
\end{table}

  \begin{figure}[htb]
   \hspace{.1\textwidth}
   \vbox{\epsfxsize=.7\textwidth
    \epsfbox{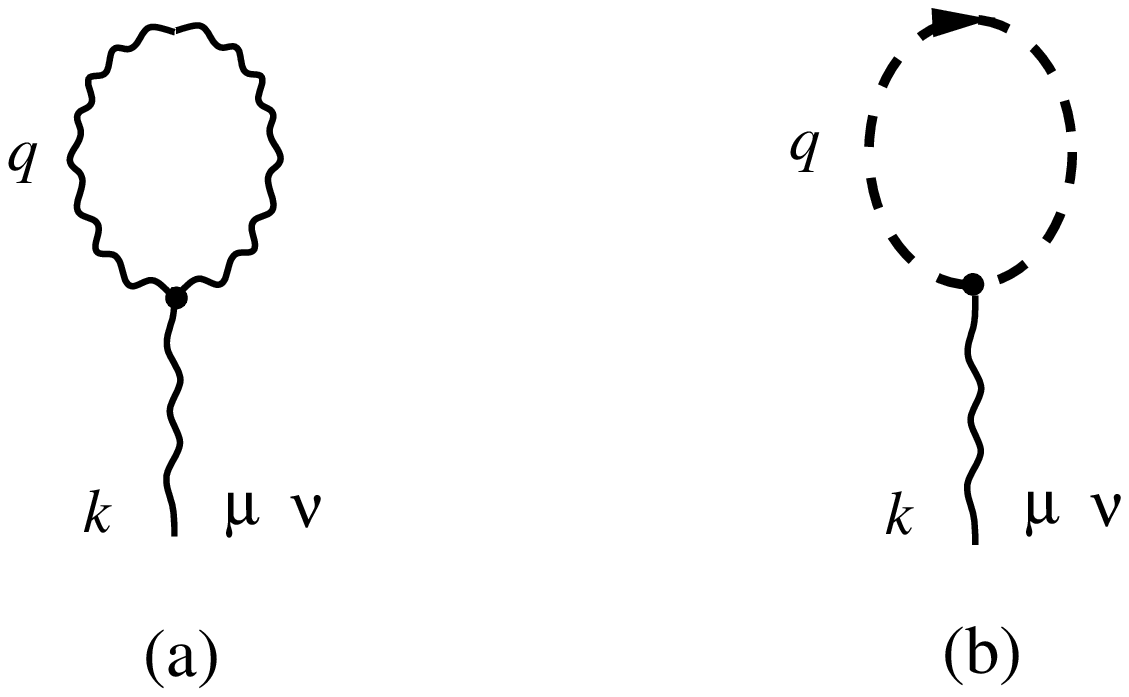}}
   \label{Fig1}
\caption{Diagrams contributing to the one-graviton function.
Wavy lines represent gravitons and dashed lines represent ghosts}
  \end{figure}
\begin{center}
  \begin{figure}[htb]
   \hspace{.1\textwidth}
\begin{center}
   \vbox{\epsfxsize=1.0\textwidth
    \epsfbox{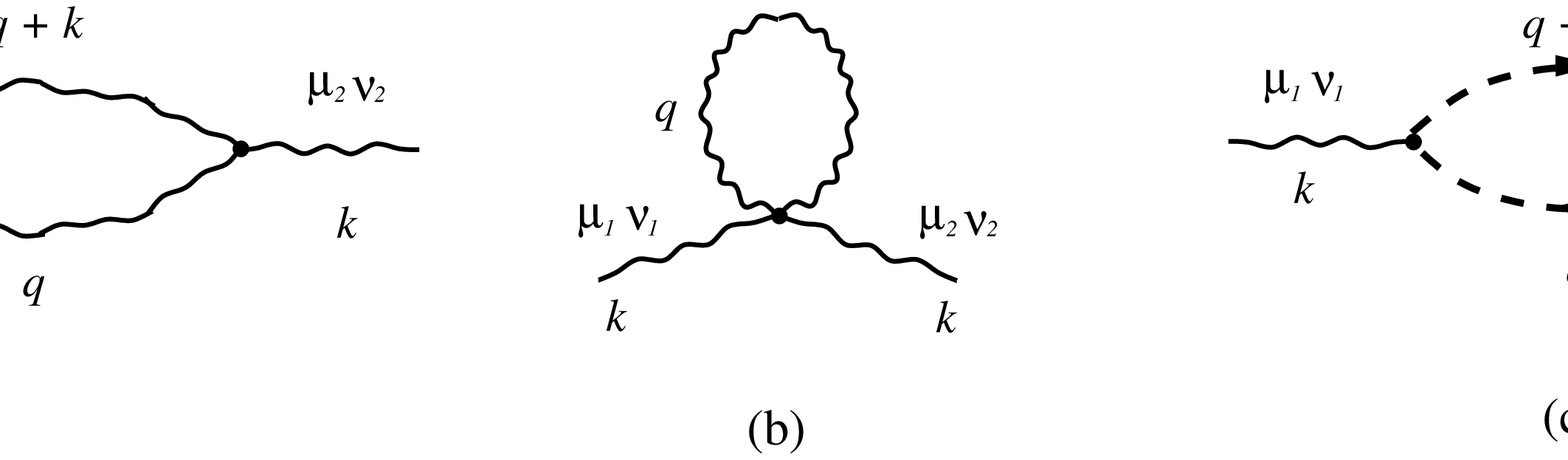}}
\end{center}
   \label{Fig2}
\caption{Diagrams contributing to the two-graviton function.
Wavy lines represent gravitons and dashed lines represent ghosts}
  \end{figure}
\end{center}
 \begin{figure}[htb]
   \hspace{0.001\textwidth}
   \hbox{\epsfxsize=.33\textwidth\epsfbox{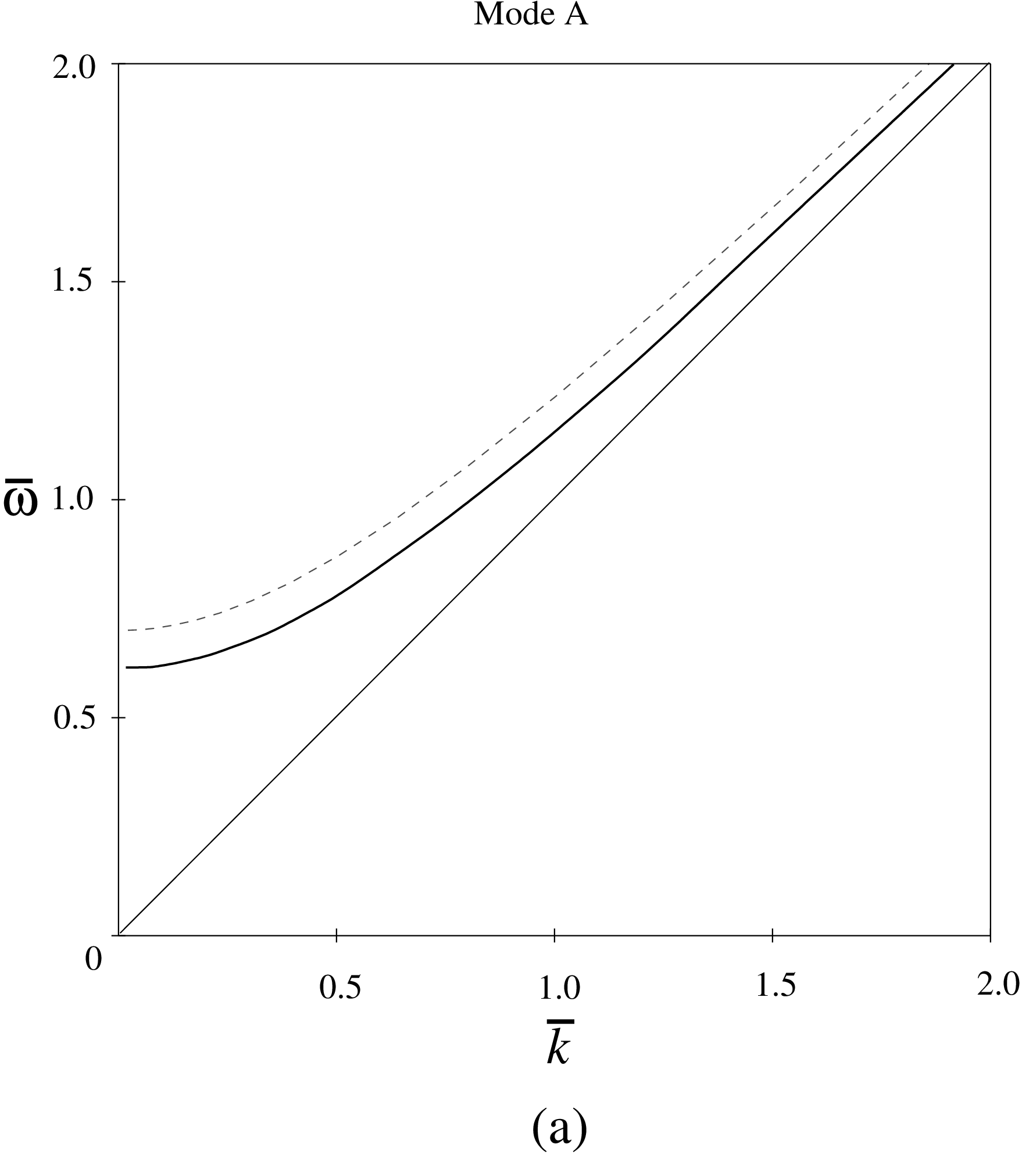}$\;\,$
         \epsfxsize=.33\textwidth\epsfbox{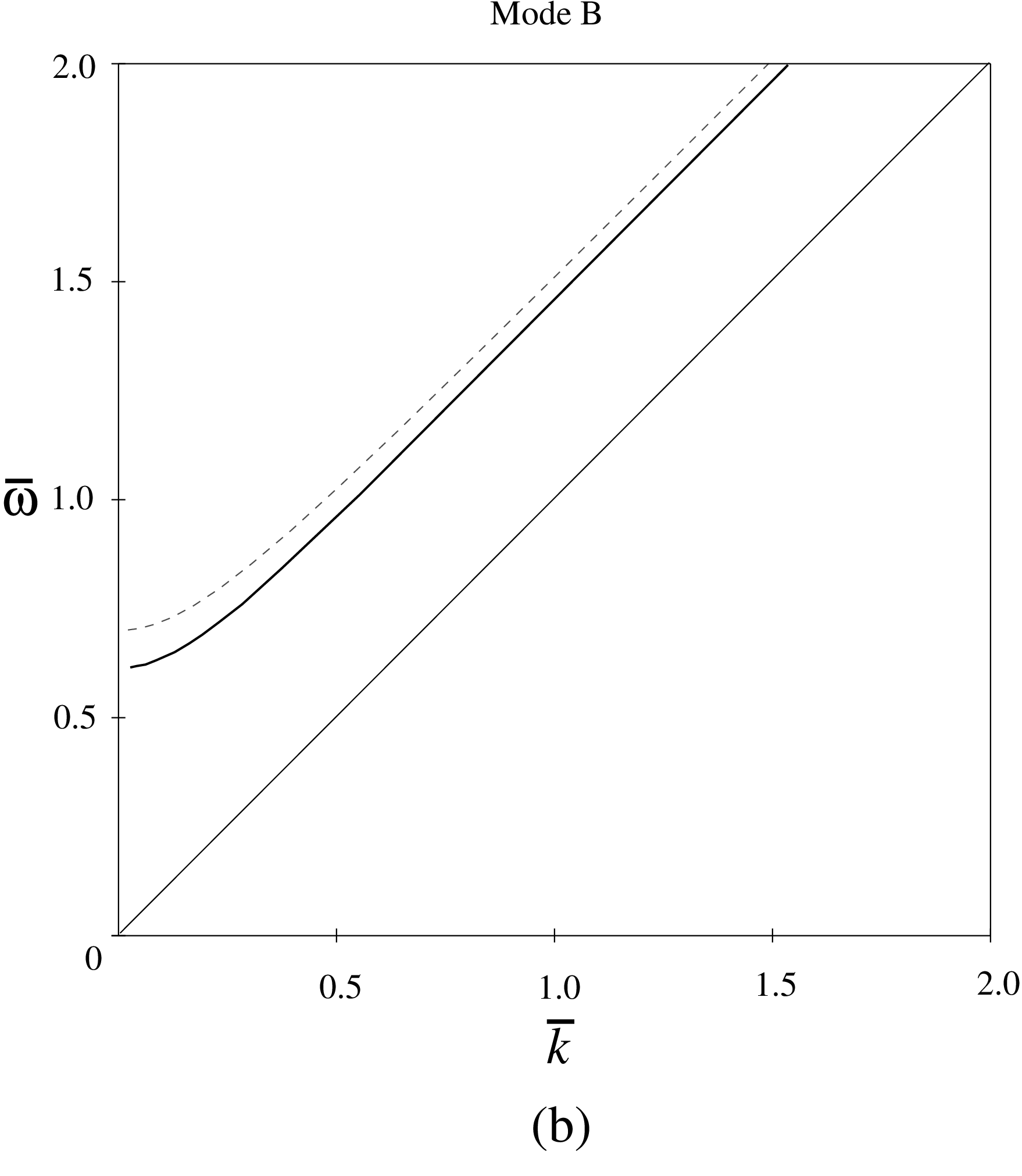}$\;\,$
         \epsfxsize=.33\textwidth\epsfbox{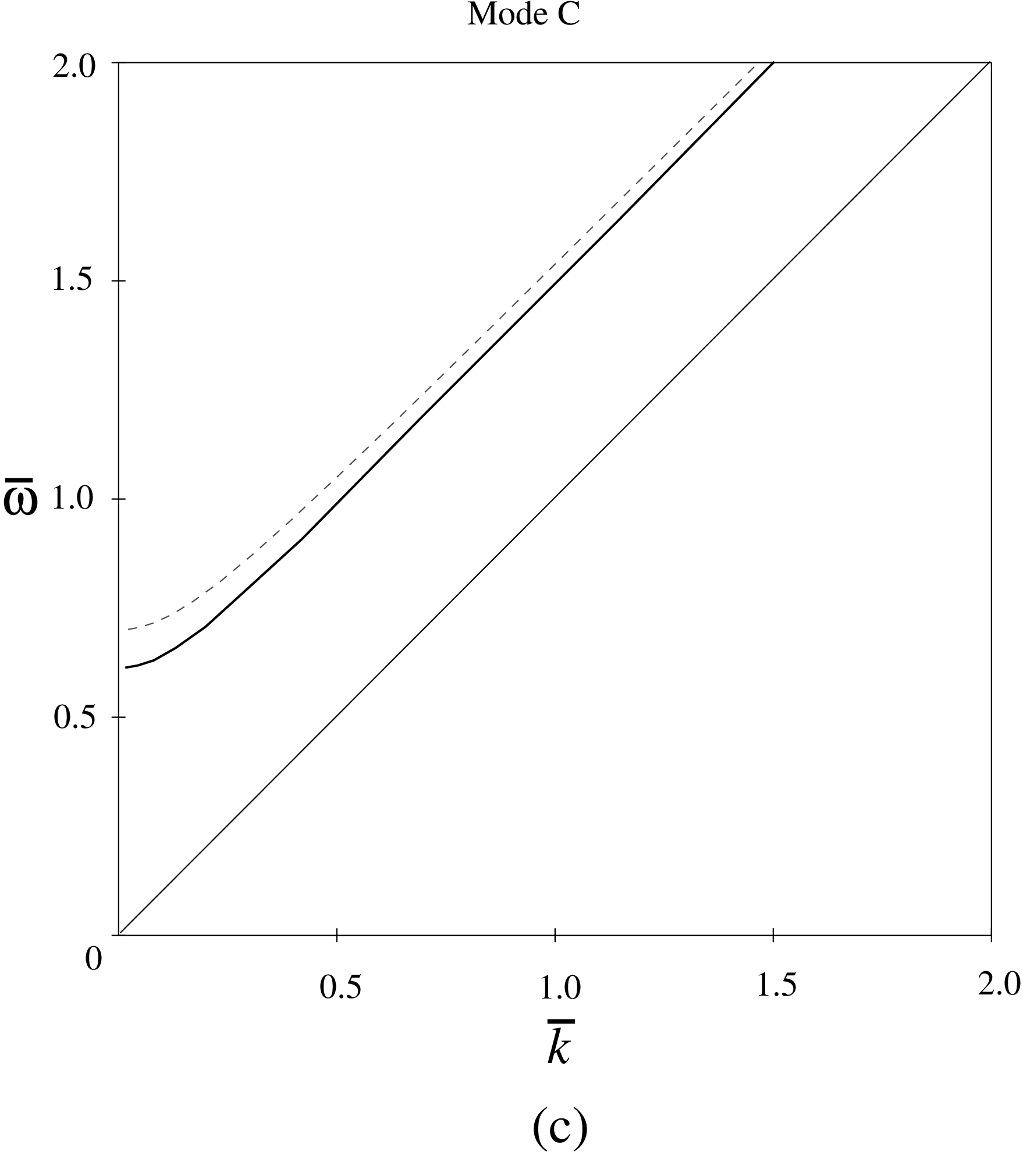}}
   \label{Fig3}
\caption{The dispersion relations for the modes A, B and C in
units of $\left(16\pi G\rho\right)^{1/2}$ 
for real frequencies and wave vectors. 
The dashed lines stand for the leading
$T^4$ contributions and the full lines represent the inclusion
of the $T^2$ sub-leading corrections for $G\, T^2=0.01$.
The light-cone is represented by the diagonal.}
  \end{figure}

\begin{figure}[htb]
   \hspace{.001\textwidth}
   \vbox{\epsfxsize=.7\textwidth
    \epsfbox{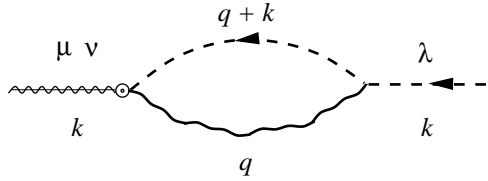}}
   \label{Fig4}
\caption{The source-ghost Feynman diagram. The full/wave line on the left represents the
external source.}
  \end{figure}
\end{document}